\documentstyle[12pt]{article}

\title{\bf Dynamical generation of space-time signature by spontaneous symmetry breaking}
\author{F. Darabi$^{a}$\thanks{e-mail:
f.darabi@azaruniv.edu} \\ $^{a}${\small Department of Physics,
Azarbaijan University of Tarbiat Moallem , Tabriz, 53714-161
Iran.} }
\begin{document}
\maketitle
\begin{abstract}
The problem of dynamical generation of 4-D space-time signature at
small scales and its stabilization towards Lorentzian signature at
large scales is studied in the context of Higgs mechanism in a
two-time scenario. It is also shown that Lorentz invariance at
small scales can be violated but at large space-time scales is
restored.
\end{abstract}
\newpage

The initial idea of signature change was due to Hartle, Hawking
and Sakharov \cite{HHS}. This idea would make it possible to have
both Euclidean and Lorentzian metrics in the path integral
approach to quantum gravity. However, it was later shown that
signature change may happen, as well, in classical general
relativity \cite{CSC}. This issue has recently been raised in the
Brane-world scenario, as well \cite{BW}. There are two different
approaches to this problem : continuous and discontinuous. In the
continuous approach, in passing from Euclidean to Lorentzian
region, the signature of the metric changes continuously, hence
the metric becomes degenerate at the border. In the discontinuous
approach, however, the metric is non-degenerate everywhere and
discontinuous at the border.

Most of the works regarding the signature change dealt with
situations where the signature changing metric is defined {\it
apriori} on the manifold and one looks for the effects of the
assumed signature change on the Einstein equations or propagation
of particles in such a manifold. However, there are some other
viewpoints in which the signature generation of the large scale
space-time is studied and considered to be a dynamical phenomenon
\cite{Pre}, \cite{Gre}, \cite{Odin}. On the other hand, it is
believed that while the signature of the large scale 4-D
space-time is definitely Lorentzian, the phenomena of dynamical
topology and signature changes at ultrashort distances can happen
as the microscopic fluctuations of the space-time. This is because
a more general formulation of gravitation should accommodate
geometries with degenerate metrics and nontrivial topologies. It
is then interesting to introduce one mechanism which accommodates
both small and large scales and addresses the problem of metric
signature at these hierarchial scales.

In the present letter, we propose such a mechanism for signature
fluctuations at ultrashort distances and stabilization of
Lorentzian signature at large scale 4-D space-time, in the context
of Higgs mechanism in a two-time scenario. We introduce a Higgs
potential whose minima will define the signature of the 4-D
metric. The parameter of the symmetry breaking is so chosen that
it leads to a quantum oscillating signature at ultrashort distance
and a definite signature at large scale. It is then discussed that
the Lorentzian signature of the present large scale universe might
have been generated due to a quantum tunneling effect at very
early universe and that the immediate inflation could have
stabilized this chosen Lorentzian metric and prevented this new
baby universe from re-tunneling to an Euclidean phase. The
subsequent Big Bang and the observed acceleration of the universe
are then considered as different ways by which the universe could
have fixed the Lorentzian metric as the preferred one.

Beside this scenario, it is also shown that at the present large
scale Lorentzian universe there can be fluctuating signature at
small scales which may be accompanied by a constant time-like
vector that accounts for a principle violation of the Lorentz
invariance at small scales. At large scales, although a same
vector can in principle exist but its norm is vanishing and so is
non-observable. This leaves the Lorentz invariance as an almost
exact symmetry at large space-time scales.

Consider the 5-D two-time metric\footnote{Such two-time metrics
are currently the subject of investigations. \cite{Wess}.}
\begin{equation}
dS_{(5)}^2=<\Phi>dt^2+ds^2-dT^2, \label{0}
\end{equation}
where $ds^2$ accounts for 3-space metric, $T$ is the extra time
dimension, and $<\Phi>$ is assumed to be the vacuum expectation
value of a dimensionless Higgs field with the following potential
\cite{leblac}
\begin{equation}
V(\Phi)=\frac{1}{2}\alpha
\Phi^2+\frac{1}{4!}\beta\Phi^4+\frac{1}{6!}\gamma\Phi^6, \label{1}
\end{equation}
where $\beta<0$ and $\gamma>0$ together with $\alpha$ are the
parameters of the potential. It is assumed that the 4-dimensional
metric is independent of the extra time, but the Higgs field
depends merely on $T$. The choice of $\alpha$ is of particular
importance in incorporating the notions of large and small scales
in the study of signature dynamics. In fact, no absolute line of
demarcation can exist between small and large scales without
having a positive definite measure of distance. Therefore, we
assume a characteristic size in the ultrashort distance regime,
described by an absolute scale of length $l_0$, which acts as a
sort of universal length that determines a lower bound on any
scale of length probed in a measurement process. The existence of
universal length $l_0$ is not compatible with the universal
requirement of Lorentz invariance. Such a violation of Lorentz
invariance may be a consequence of unification of quantum physics
and gravity and is expected to manifest itself at ultrashort
distances. We therefore take $\alpha= l_0/l$ where $l$ is the
characteristic size of the region over which one measures the
signature of metric.

By starting from a very large parameter $\alpha\gg1$, namely $l\ll
l_0$, one finds that the potential has one minimum at $<\Phi>=0$
as is shown in Fig.1. However, at some ( critical ) smaller
parameter $l=l_c<l_0$ ( $l_c$ is presumably the Planck length
$l_p$ ), the potential will have three minima at points where
$V(\Phi)=0$ ( see Fig.2 ). One of these minima is at $<\Phi>=0$
and two others are at $\pm<\Phi>_c$. In other words, at $l=l_c$
there are two phases in equilibrium with each other, one with
$<\Phi>=0$ and the other with $<\Phi>_c$ ( or $-<\Phi>_c$ ). The
phase with $<\Phi>=0$ is stable when $l \leq l_c$, and {\it
meta-stable } when $l$ is a bit larger than $l_c$. The order
parameter, namely $<\Phi>$, is discontinuous at the transition
$l=l_c$. Therefore, we are dealing with a {\it first order phase
transition} where two phases can coexist, one with $<\Phi>=0$ and
the other with nonzero $<\Phi>$. One may then find meta-stability
so that the system can persist in the phase with $<\Phi>=0$, for
the parameter $l$ very close to $l_c$. When $l=l_0$, the potential
$V(\Phi)$ is unstable at $<\Phi>=0$, but has two negative minima
which are stable, as Fig.3. In such case, there can never be
coexistent phases at $<\Phi>=0$ and $<\Phi>\neq 0$. Therefore, the
unstable phase at $<\Phi>=0$ is removed and a stable phase should
be chosen out of two minima at $\pm <\Phi>_0$.

This means, when $l\ll l_0$ we have the degenerate 4-D metric
\begin{equation}
dS_{(4)}^2=ds^2,
\end{equation}
everywhere on the manifold, before symmetry breaking. Then, at
some smaller parameter $\alpha$ namely $l=l_c<l_0$ the 4-D metric
is capable of taking on the following forms
\begin{equation}
\left \{ \begin{array}{ll} dS_{(4)}^2=<\Phi>_c
dt^2+ds^2,\:\:\:\:\:\:\:\:\:\:\: Euclidean
\\
dS_{(4)}^2=ds^2,\:\:\:\:\:\:\:\:\:\:\:\:\:\:\:\:\:\:\:\:\:\:\:\:\:\:\:\:\:\:\:\:\:\:\:\:
Degenerate
\\
dS_{(4)}^2=-<\Phi>_c dt^2+ds^2,\:\:\:\: Lorentzian
\end{array}\right.
\label{2}
\end{equation}
corresponding to three equal minima of the potential where
$V(\Phi)=0$, so that they can be in equilibrium with each other.
The manifold at the scale $l_c$ is then capable of being
degenerate, Euclidean, and Lorentzian. In other words, the
signature of metric oscillates ( in $T$ ) between the above three
forms due to quantum tunneling between the minima of the
potential. If the parameter $l$ begins to increase above $l_c$ the
system will provisionally be meta-stable so that the metric can
persist, for the parameter $l$ close to $l_c$, in coexistent
phases. Finally, at $l=l_0$ the system is unstable at $<\Phi>=0$
and has two negative minima. This means, the manifold can no
longer have a degenerate metric and the signature can just
oscillate between Euclidean and Lorentzian, due to quantum
tunneling between the two minima. As $l$ increases above $l_0$ the
system becomes more stable so that for $l\gg l_0$ the tunneling
probability approaches zero and the preferred signature is
permanently fixed.

A few words on the length scales $l_0$ and $l$ are in order. In
one hand, the existence of a universal length $l_0$ in the small
scale regime is in sharp contrast with the universal requirement
of Lorentz invariance. On the other hand, one can not distinguish
between large and small scales without having a positive definite
metric. According to Blokhintsev \cite{Blok}, accompanying the
notion of a universal length ( Lorentz non-invariance ) with a
constant time-like vector $N_{\mu}=(1, 0, 0, 0)$, it is possible
to distinguish between small and large scales in Minkowski
space-time by taking the positive definite interval
\begin{equation}
S_{(4)}^2=\bar{\eta}_{\mu \nu}x^{\mu}x^{\nu},\:\:\:
\bar{\eta}_{\mu \nu}={\eta}_{\mu \nu}+2N_{\mu}N_{\nu}.
\end{equation}
Therefore, the existence of a universal length $l_0$ at ultrashort
distance regime is inevitably related to the existence of an {\it
internal} time-like vector $N_{\mu}$ over the manifold. Given the
Euclidean metric $\bar{\eta}_{\mu \nu}$ one may determine the
absolute size of a distance by comparing $S_{(4)}$ with the
universal length $l_0$. In the same way, one may determine the
absolute size of the distance $l$ and determine the meaningful
value of the parameter $\alpha$.

With no loss of generality one may take the distance $l$ to be the
radius of universe, namely the scale factor $R$. Therefore, one
may expect different phases in the potential according to the
evolution of $R$. At very early universe $R\ll l_0$ ( or $R< l_c$
) we have not a meaningful notion of the metric because it is
degenerate. At $R=l_c<l_0$, the metric can fluctuate between
Euclidean, degenerate and Lorentzian forms. This is plausible in
the quantum gravity regime $l_c=l_p$ and is consistent with
quantum tunneling in cosmology in which the universe tunnels from
nothing ($R=0$), through a Euclidean region, to the planck length
$l_p$ at which the universe may fluctuate between Euclidean
($R\leq l_p$), degenerate ($R= l_p$) and Lorentzian ($R\geq l_p$)
forms. As long as the scale factor is close to the planck length
these quantum oscillations ( in $T$ ) persist with no preference
for Euclidean or Lorentzian metric to be the permanent one. This
means the true vacuum of the potential is not yet selected.
However, once some suitable initial conditions are provided for
one of the quantum tunnelings to the Lorentzian region, the
universe can start time ( $t$ ) evolution in the scale factor
towards $l_0$ and this stabilizes the Lorentzian metric because
$\alpha$ decreases. Therefore, at the scale $R=l_0>l_p$ the metric
is definitely Lorentzian which means $-<\Phi>_0$ is selected as
the preferred minimum of the potential. However, the probability
for quantum tunneling to the Euclidean region is not yet excluded,
because the barrier between the two minima is not so high and
wide. The new born Lorentzian universe can get rid of death (
thorough quantum re-tunneling to the Euclidean region ) by
undergoing an inflation in the scale factor. This inflation
launches the small scale factor $R=l_0$ ( presumably the Grand
unification scale ) to a distance tens of order greater than $l_0$
in a very short period of time, namely $10^{-35}-10^{-33}$
seconds. Hence $\alpha$ becomes very small and leads to a very
high barrier so that the probability for quantum tunneling becomes
very small, as well. In this regard, the inflation helps the
Lorentzian metric to be stabilized in a tiny fraction of a second.
After inflation, the big bang also causes the universe to be
expanded rapidly which leads to more stabilization of the
Lorentzian metric. This stabilization is continued as long as the
universe expands. One may then interpret the expansion of the
universe as a way to avoid re-tunneling to the Euclidaen metric.
The recent observed acceleration of the universe may also be
addressed and justified according to this scenario.

Now suppose there is a constant internal vector on the 4-D
manifold as
\begin{equation}
N_{\mu}=(\sqrt{\alpha|<\Phi>|}, 0, 0, 0), \label{3}
\end{equation}
associated with the properties of vacuum through $<\Phi>$. Then,
for $R\ll l_0, \alpha \gg 1$ we have $N_{\mu}=(0, 0, 0, 0)$
corresponding to the absolute minimum at $<\Phi>=0$, before
symmetry breaking. When $R=l_c$, the parameter $\alpha$ decreases
and the vector $N_{\mu}$ is found in a non-vanishing form, namely
$N_{\mu}=(\sqrt{\alpha|\pm<\Phi>_c|}, 0, 0, 0)$. This means that
before symmetry breaking, when $R \ll l_0$, there is no $N_{\mu}$
at all, and one is appeared at the symmetry breaking at $R=l_c$.
At $R >l_c$ where only Euclidean and Lorentzian metrics are
available through the quantum tunneling, one may use this
non-vanishing $N_{\mu}$ to relate the Euclidean metric
$\bar{g}_{\mu \nu}$ and Lorentzian one $g_{\mu \nu}$
\begin{equation}
\bar{g}_{\mu \nu}=g_{\mu
\nu}+\frac{2}{\alpha}N_{\mu}N_{\nu},\label{4}
\end{equation}
with no restriction on the norm of $N_{\mu}$. This is because the
permanent metric is not fixed. However, far from the critical
point, $R \gg l_c$ where the true vacuum of the system is singled
out as the Lorentzian metric $g_{\mu \nu}$, we find, by
contracting Eq.(\ref{4}) with ${g}^{\mu \nu}$, that the internal
vector has negative norm
$$
N_\mu N^\mu=-\alpha,
$$
which accounts for the time-like nature of $N_{\mu}$, as is
required according to Blokhintsev point of view. Notice that the
norm of $N_{\mu}$ is almost vanishing for $\alpha\ll 1$, namely
$R\gg l_0$, and becomes important at small scales $R\simeq l_0$.
Therefore, the internal vector $N_{\mu}$ is non-observable at
large scales which means the Lorentz invariance is an almost exact
symmetry at large distances compared with $l_0$.

\section*{Concluding remarks}

We have discussed that as long as suitable initial conditions for
time evolution of the scale factor were not satisfied at very
early universe the 4-D metric could have been oscillating quantum
mechanically between Euclidean and Lorentzian phases as degenerate
vacuums of a Higgs potential. However, once these conditions were
satisfied in one of the tunnelings into a Lorentzian vacuum (
early universe ), the immediate inflation and the subsequent Big
Bang could have stabilized this chosen metric . In this regard,
inflation, Big Bang and even the present observed acceleration of
the universe can be considered as different behaviors of the
universe to escape the death through re-tunneling to the Euclidean
metric. We have also shown that Lorentz invariance at small scales
can be violated but at large space-time scales it can be restored
exactly.

\section*{Acknowledgment} This work has been supported financially
by the Research Department of Azarbaijan University of Tarbiat
Moallem, Tabriz, Iran.\\
\newpage
{\large{\bf Figure captions}}
\vspace{10mm}\\
Figure 1. The potential at $l\ll l_0$.\\
\\
Figure 2. The potential at $l=l_c<l_0$\\
\\
Figure 3. The potential at $l=l_0$ \\

\end{document}